\documentclass[12pt]{iopart}
\usepackage{graphicx}

\usepackage{iopams}

\begin{document}

\noindent
{\bf Biological modelling and information}

\title[Biological molecular machines can process information]{Biological molecular machines can process information to reduce energy losses}

\author{Michal Kurzynski and Przemyslaw Chelminiak}

\address{Faculty of Physics, A. Mickiewicz University, Umultowska 85, 61-614 Poznan, Poland}
\ead{kurzphys@amu.edu.pl}
\vspace{10pt}
\begin{indented}
\item[]December 2018
\end{indented}

\begin{abstract}
Biological molecular machines are enzymes that simultaneously catalyze two processes, one donating free energy and second accepting it. Recent studies show that most native protein enzymes have a rich stochastic dynamics that often manifests in fluctuating rates of the catalyzed processes and the presence of short-term memory resulting from transient non-ergodicity. For such dynamics, we prove the generalized fluctuation theorem predicting a possible reduction of energy dissipation at the expense of creating some information stored in memory. The theoretical relationships are verified in computer simulations of random walk on a model critical complex network. The transient utilization of memory may turn out to be crucial for the movement of protein motors and the reason for most protein machines to operate as dimers or higher organized assemblies. From a broader physical point of view, the division of free energy into the operation and organization energy is worth emphasizing. Information can be assigned a physical meaning of  a change in the value of both these functions of state.
\end{abstract}

\begin{indented}
\item {\bf Keywords:} single molecule, molecular machines, complex networks, information processing, generalized fluctuation theorem, Maxwell's demon
\end{indented}

%
%
\maketitle
%
%

\section{Introduction}

The basic task of statistical physics is to combine the dynamics of the  microstates of a studied system with the dynamics of the macrostates. The transition from the microstates to macrostates consists on averaging over time \cite{Penr79}. In the statistical physics of simple systems, there are no intermediate levels of organization between microscopic mechanics and macroscopic thermodynamics. Living matter is, however, a complex system with the entire hierarchy of organization levels \cite{Volk09,Eign13}. The new achievements at the turn of the century allowed us to understand more accurately the nature of the micro and macrostates of the biological molecular machines, which belong to the level currently being referred to as nanoscopic. 

First, the tertiary structure of the enzymatic proteins, formerly identified with a particular conformational state, has been extended to the whole network of conformational substates \cite{Karp02,Frau10,Kita98,Kurz98,Lu98,Engl06,Henz07,Uver10,Chod14,Shuk15}. The Markov stochastic process on such a network is to be considered as the starting microscopic dynamics for a statistical treatment. The rule are fluctuating rates of the catalyzed reactions (the ``dynamic disorder") \cite{Lu98,Engl06,Kurz08}, which means realization of them in various randomly selected ways. As a symbol of the progress made recently, the molecular dynamics study of conformational transitions in the native phosphoglycerate kinase could be quoted, in which a network of 530 nodes was found in the long, 17~$\mu$s simulation \cite{Hu16}. This network seems to be scale-free and display a transition from the fractal to small-world organization \cite{Roze10,Esco12}. On such networks, modeled by scale-free fractal trees \cite{Goh06} extended by long-range shortcuts, the random choice of  reaction realization is quite natural \cite{Kurz14}.

Secondly, new methods of stochastic thermodynamics have been applied to the description of the non-equilibrium behavior of single nanoobjects in a finite time perspective. Work, dissipation and heat in nanoscopic systems are random variables and their fluctuations, proceeding forward and backward in time, proved to be related to each other by the fluctuation theorem \cite{Evan02,Jarz97,Croo99,Seif12}. For nanoscopic machines, the notion of information creation can be defined and the relationships between entropy production and information lead to the generalized fluctuation theorem  \cite{Saga10,Ponm10,Saga12,Mand12,Bara14,Hart14,Horo14,Shir15,Shir16}. It strengthened an almost universal consensus regarding the view on the operation of  Maxwell's demon consistent with thermodynamics \cite{Mand12,Luma14,Maru09,Saga13,Parr15}. Thus, the demon must have a memory and in order to use fluctuations to do the work, it reduces entropy at the expense of gathering the necessary information on fluctuations in this memory. The information must be erased sooner or later, and for this the same or greater work must be used.

The purpose of this paper is to answer the question of whether the particular stochastic dynamics of the biological molecular machines allows them to act as Maxwell's demons. This would help to understand the hard-to-interpret behavior of single-headed biological motors: myosin II \cite{Kita99}, myosin V \cite{Wata04}, kinesin 3 \cite{Okad99}, and both flagellar \cite{Koji02} and cytoplasmic \cite{Mall04} dynein, which can take more than one step along their track per ATP or GTP molecule hydrolyzed. Basing on recent investigations, we assume that the stochastic dynamics of the enzymatic machine is characterized by transient memory and the possibility to realize the catalyzed process in various randomly selected ways. For such dynamics we calculate both the energy dissipation and information, and show that they can be of the opposite sign, hence the first can be partially compensated by the second. Because available experimental data on the actual conformational transition networks in native proteins is still very poor, we verify the theoretical results only by computer simulations for a model network. Our conclusions are based on analysis of the simulated time course of the catalyzed processes expressed by the strings of discrete jumps at random moments of time. Since similar signals can be registered in the experiments \cite{Kita99,Wata04,Okad99,Koji02,Mall04}, all the theses of the paper are open for experimental verification. Thus, besides presentation of new theoretical concepts, the paper can be also treated as an invitation to experimentalists to perform similar analysis on real systems.

\section{Clarification of concepts}

\subsection*{\bf 2.1. Biological molecular machines as nanoscopic chemo-chemical machines}

\begin{figure}[ht]
	\hspace{6pc}
	\centering
	\includegraphics[scale=0.8]{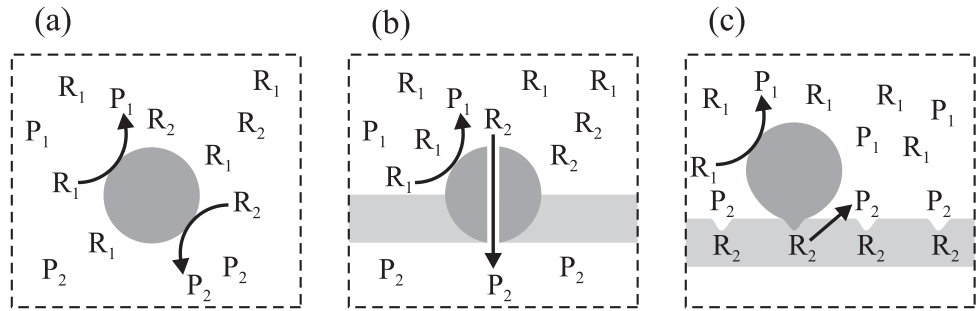}
	\caption{A schematic picture of the three types of the biological molecular machines. (a)~Enzymes that simultaneously catalyze two reactions. (b)~Pumps placed in a biological membrane -- the molecules present on either side of the membrane can be considered to occupy different chemical states. (c)~Motors moving along a track -- the external load influences the free energy of motor binding, which can be considered as a change in the probability of finding a particular track's binding site, hence its effective concentration \cite{Kurz06,Kolo07}. Constraints, symbolized by a frame of a broken line, are imagined by the observer to control the number of the externally incoming and outgoing molecules involved in the catalysis. It should be stressed that the entire system inside the frame represents the machine: 
	it is these constraints that determine the values of the thermodynamic variables and the conjugate forces acting on the machine.}
	\label{fig1}
\end{figure}

For many historical reasons, the word ``machine" has aquired many different meanings. 
In the thermodynamic context, a machine is understood to be a physical system that mediates in performing work between two other systems related to it. Work can be done by mechanical, electrical, chemical, thermal or still some other forces. The forces are defined as the differences of the respective potentials \cite{Call85}. Because the biological molecular machines operate using thermal fluctuations, just like chemical reactions, we treat them as chemo-chemical machines \cite{Kurz06}. The protein chemo-chemical machines are to be considered as enzymes, that simultaneously catalyze two effectively unimolecular chemical reactions: the energy-donating input reaction ${\rm R}_1 \rightarrow {\rm P}_1$ and the energy-accepting output reaction ${\rm R}_2 \rightarrow {\rm P}_2$, see figure~\ref{fig1}(a). Also pumps and molecular motors can be treated in the same way, see figures~\ref{fig1}(b) and (c). 

The chemo-chemical machine we consider consists of a single enzyme macromolecule, surrounded by a solution of its substrates and products (figure~\ref{fig1}). The whole is an open non-equilibrium system which operates stochastically under the influence of forces controlled by the concentrations of reactants. Under specified relations between the concentrations of all reactants \cite{Kurz14}, the non-equilibrium molar concentrations of product molecules [P$_1$] and [P$_2$], related to enzyme total concentration [E], are the two dimensionless thermodynamic variables $X_1$ and $X_2$ determining the state of the system. These, together with the conjugate thermodynamic forces, i.e. chemical affinities (the differences in chemical potentials) $A_1$ and $A_2$, determine work performed on and by the machine, respectively. Assuming that the molecule solution is perfect, the formal definitions have the form \cite{Kurz06,Hill89}:
\begin{equation}
\label{eq1}
X_{i}:=\frac{[\mathrm{P}_{i}]}{[\mathrm{E}]}\,,\hspace{4pc}
\beta A_{i}:=\ln\frac{[\mathrm{P}_{i}]^{\mathrm{eq}}}{[\mathrm{R}_{i}]
	^{\mathrm{eq}}}\frac{[\mathrm{R}_{i}]}{[\mathrm{P}_{i}]}\,.
\end{equation}
$\beta$ is the reciprocal of the thermal energy $k_{\rm B}T$ and the superscript eq denotes the  equilibrium concentrations. 

The observable action of the machine is its turnovers under the influence of forces $A_1$ and $A_2$ measured by the number of molecules P$_1$ and P$_2$ created in a unit of time. Establishing the forces $A_1$ and $A_2$ and the variables $X_1$ and $X_2$ defining them is formally equivalent to holding the machine in the steady-state conditions due to imagined constraints that control the number of incoming and outgoing reacting molecules, what is symbolically illustrated in figure~\ref{fig1}.

\subsection*{\bf 2.2. Energy processing in stationary isothermal machines}

\begin{figure}[t]
	\vspace{1pc}
	\hspace{4pc}	
	\centering
	\includegraphics[scale=0.88]{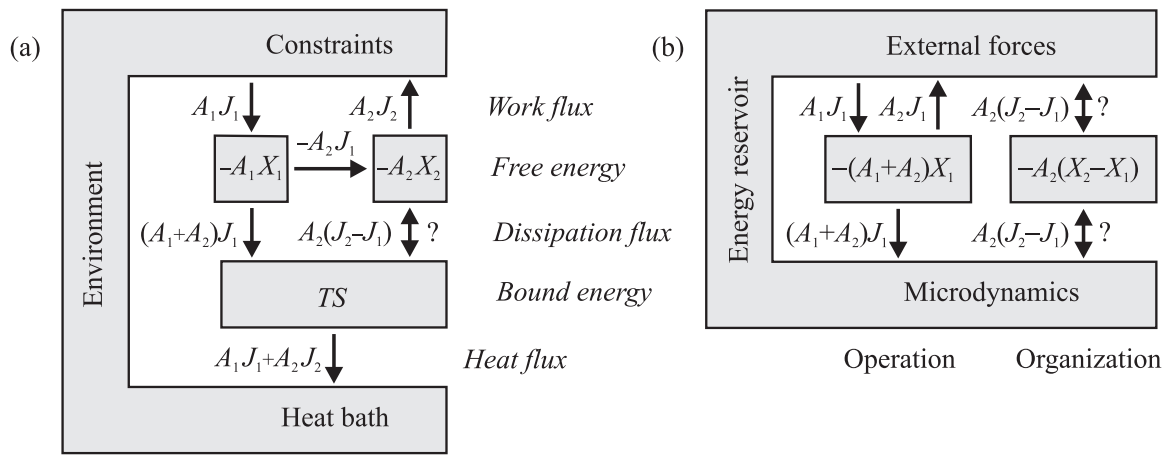} 
	\caption{Energy processing in the stationary isothermal machine. The constraints keep stationary values of thermodynamic variables $X_1$ and $X_2$ fixed. (a)~Division of the machine's internal energy into free energy $F = -A_1X_1-A_2X_2$ and bound energy $TS$. The directions of the energy fluxes shown are for $J_1, J_2, A_1 > 0$ and $A_2 < 0$. The direction of flux $A_2(J_2-J_1)$, marked with a question mark, is a subject of discussion in this paper. (b)~The alternative view of energy processing in the stationary isothermal machine, which is described in the text. Only free energy, observable macroscopically, is specified. The bound energy, determined by unknown microdynamics, is considered, along with the heat bath and imagined constraints, as the energy reservoir.}
	\label{fig2}
\end{figure}

The biological molecular machines operate at a constant temperature. Under isothermal conditions, the internal energy is uniquely divided into free energy, the component that can be turned into work, and entropy multiplied by temperature (after Helmholtz, we call it bound energy), the component that can be turned into heat \cite{Call85}. Both thermodynamic quantities can make sense in the non-equilibrium state if the latter is treated as a partial equilibrium state \cite{Kurz06}. Free energy can be irreversibly transformed into bound energy in the process of internal entropy production which, in the energy-related context, means energy dissipation. In view of such internal energy division, the protein molecular machines are referred to as free energy transducers \cite{Hill89}. The energy processing pathways in a stationary isothermal machine are shown in figure~\ref{fig2}(a), where the role of all the physical quantities being in use is also indicated. $X_i$ denotes the input ($i=1$) or the output ($i=2$) thermodynamic variable, $A_i$ is the conjugate thermodynamic force and the time derivative, $J_i = {\rm d}X_i/{\rm d}t$, is the corresponding flux. $T$ is the temperature and $S$ is the entropy. To clearly specify the degree of coupling of the fluxes, $\epsilon := J_2/J_1$, it is important that variables $X_1$ and $X_2$ be dimensionless as in equation (\ref{eq1}). Corresponding forces $A_1$ and $A_2$ are then also dimensionless, if only multiplied by the reversal of the thermal energy $\beta = (k_{\rm B}T)^{-1}$. We assume $J_1, J_2, A_1 > 0$ and $A_2 < 0$ throughout this paper but  $-A_2 \leq A_1$, i.e., variables $X_1$ and $X_2$ are defined such that the machine does not work as a gear.  

In the steady state, the total work flux (the processed power) $A_1J_1 + A_2J_2$ equals the heat flux and both equal the dissipation flux (the rate of internal entropy production multiplied by the temperature), which, according to the second law of thermodynamics, must be nonnegative. However, it consists of two components. The first component, $(A_1+A_2)J_1$, achieved when the fluxes are tightly coupled, $J_2=J_1$, must also be, according to the same law, nonnegative, but the sign of the complement $A_2(J_2-J_1)$ is open to discussion. In macroscopic machines,  the latter is also nonnegative and has the obvious interpretation of a frictional slippage in the case of mechanical machines, a short-circuit in the case of electrical machines, or a leakage in the case of pumps \cite{Kurz06}. However, because of non-vanishing correlations \cite{Saga10,Ponm10,Saga12,Mand12,Bara14,Hart14,Horo14,Shir15,Shir16}, entropy $S$ in the nanoscopic machines is not additive as in the macroscopic machines, which could cause a change the sign of $A_2(J_2-J_1)$ to negative, i.e., for $A_2 < 0$, to $J_2 \geq J_1$. In fact, the ratio $\epsilon = J_2/J_1$ for the single-headed biological motors mentioned in the Introduction was observed to be higher than unity \cite{Kita99,Wata04,Okad99,Koji02,Mall04}.

From the point of view of the output force $A_2$, subsystem 1 carries out work on subsystem 2 while subsystem 2 carries out work on the environment. Jointly, the flux of the resultant work (the resultant power) $A_2(J_2-J_1)$ is driven by the force $A_2$. The complement to $A_1J_1+A_2J_2$ is flux $(A_1+A_2)J_1$ driven by the force $A_1+A_2$. Consequently, the free energy processing from figure~\ref{fig2}(a) can be alternatively presented as in figure~\ref{fig2}(b), with the free energy transduction path absent. Here, two interacting subsystems of the machine characterized by variables $X_1$ and $X_2$ have been replaced by two non-interacting subsystems characterized by variable $X_1$ and the difference $X_2-X_1$. The first describes the perfect, tightly coupled operation of the machine while the other is its deteriorating or improving organization. Therefore,  the variable $X_1$ will be referred to as operation of the machine, and the variable $X_2-X_1 $ as organization of the machine. As a result of interaction with microdynamics, both newly defined subsystems, although directly non-interacting, still remain statistically correlated.

\subsection*{\bf 2.3. Dynamics of biological molecular machines}

\begin{figure}[t]	
	\hspace{6pc}
	\includegraphics[scale=0.95]{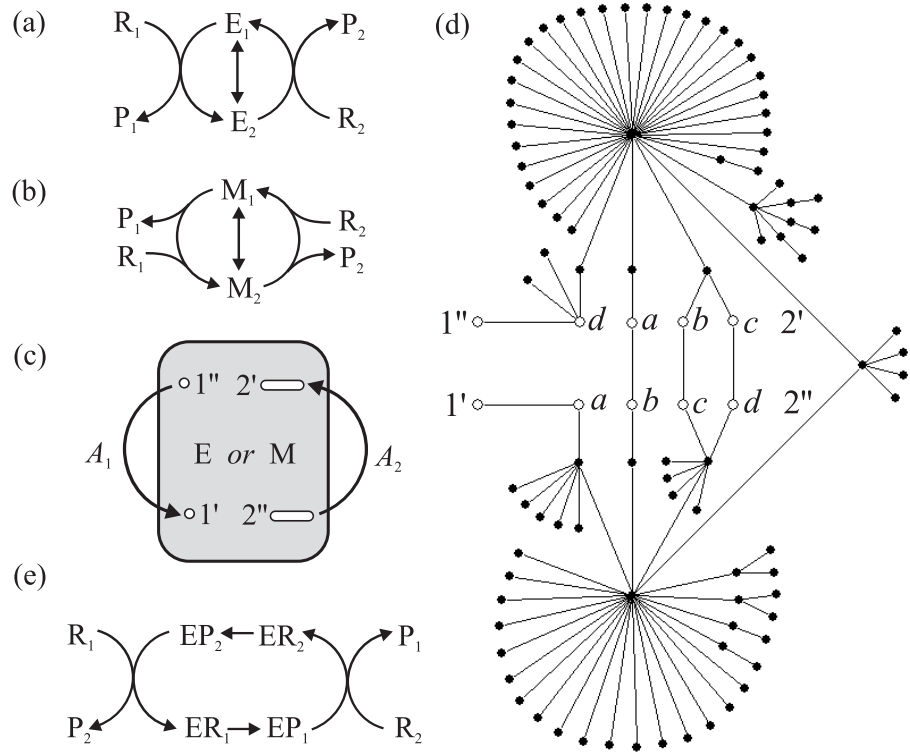}   
	\caption{Dynamics of the enzymatic chemo-chemical machine. (a) and (b)~The schemes of  conventional chemical kinetics that take into account only two states E$_1$ and E$_2$ of the free enzyme or the two states M$_1$ and M$_2$ of the enzyme-reagents complex. (c)~The generalized scheme \cite{Kurz14,Kurz03}. The gray box represents a one-component network of transitions between conformational substates that make up the enzyme or the enzyme-reactants complex native state. All these transitions satisfy the detailed balance condition. A single pair (the ``gate") of conformational states $1''$ and $1'$ is distinguished, between which input reaction ${\rm R}_1 \rightarrow {\rm P}_1$ breaks the detailed balance. Also, a single or variety (ovals) of pairs of conformational states $2''$ and $2'$ is distinguished, between which output reaction ${\rm R}_2 \rightarrow {\rm P}_2$ does the same. All the reactions are reversible; the arrows indicate the directions assumed to be forward. (d)~The sample implementation of the 100-node network, constructed following the algorithm described in appendix A. A single output gate chosen for the simulations is created by the pair of transition states $(2''a, 2'd)$. The four pairs of transition states $(2''a, 2'a)$, $(2''b, 2'b)$, $(2''c, 2'c)$ and $(2''d, 2'd)$ lying tendentiously one after another, create alternative fourfold output gate. (e)~The case of alternating reactions of reactants binding and detaching that leads to microdynamics described by the two-component network of conformational transitions separately in the complexes enzyme-reactant 1 and enzyme-reactant 2.}
	\label{fig3}
\end{figure}

In the language of conventional chemical kinetics, the action of the chemo-chemical machine can be presented as in figure~\ref{fig3}(a). The energy-donating reaction ${\rm R}_1 \rightarrow {\rm P}_1$ forces the direction of the  energy-accepting reaction ${\rm R}_2 \rightarrow {\rm P}_2$, though the nonequilibrium concentration values [R$_2$] and [P$_2$] would prefer the opposite direction \cite{Kurz06,Hill89}.
We assumed that the binding of reactants precedes their detaching but this order may be reversed as shown in figure~\ref{fig3}(b). For the sake of simplicity, we do not take into account details of the binding and detaching reactions involving spatial diffusion of reactants and molecular recognition, which should be a topic for a separate consideration \cite{Kurz03}. 

Only two states E$_1$ and E$_2$ of the free enzyme are distinguished in the scheme (a) and only two states M$_1$ and M$_2$ of  the enzyme-reagents complex are distinguished in the scheme (b). Only one non-reactive transition between them is assumed. However, as mentioned in the Introduction, the nanoscopic kinetics of a biological machine is described by the whole network of conformational substates. In this network, with transitions that obey the detailed balance condition, a system of pairs of nodes (the ``gates") is distinguished, between which the input and output chemical reactions force additional transitions that break the detailed balance in a degree dependent on the values of $A_1$ and $A_2$  \cite{Kurz14,Kurz03}, see figure~\ref{fig3}(c). The rule is the presence of many gates allowing the choice of the way of implementing reactions  \cite{Lu98,Engl06}. As mentioned, the network is prabably scale-free and display a transition from the fractal to small-world organization \cite{Hu16,Roze10,Esco12,Goh06,Kurz14}.  

A sample network of 100 nodes with such property, constructed following the algorithm described in appendix A, is depicted in figure~\ref{fig3}(d). Let us note two hubs, the substates of the lowest microscopic free energy, e.g. ``open" and ``closed", or ``bent" and ``straight", usually the only ones occupied sufficiently high to be observable. So, the proposed kind of stochastic dynamics combines in fact a chemical description with a mechanical one \cite{Kurz06}. The dream of structural biologists, increasingly fulfilled \cite{Kode10}, is to find the largest possible number of such distinguished states for actual protein machines, but let us remember that in contrast to macroscopic mechanical machines, the transitions between such states in nanomachines are always stochastic. 

If the reactions of binding and detaching reagents occur alternately, as in figure~\ref{fig3}(e), the network of conformation transitions is two-component, what can be important for the operation of the machine and is commented in chapter 5.

\section{Generalized fluctuation theorem for biological molecular machines}

In the nanoscopic systems operating under the influence of controlled forces, work, dissipation and heat in a finite time perspective are random variables and their nonequilibrium behavior is described by the fluctuation theorem \cite{Evan02,Jarz97,Croo99,Seif12}. The microscopic dynamics of the biological nanomachine is represented by the Markov stochastic process on a network to which external forces add several transitions that break the detailed balance, see figure~\ref{fig3}(c).  The partial entropy production in the distinguished set of transitions satisfies the fluctuation theorem in the form of the Jarzynski equality \cite{Shir15,Shir16}: 
\begin{equation}
\label{eq2}
\langle\exp(-\sigma)\rangle = 1.
\end{equation}
The stochastic dimensionless entropy production $\sigma$ in the external transitions,
multiplied by the thermal energy  $k_{\rm B}T = \beta^{-1}$, determines free energy dissipation for a period of time $t$ in the stationary energy processing presented in figure~\ref{fig2}, hence
\begin{equation}
\label{eq3}
\sigma = \beta[A_1\mathcal{J}_1(t) + A_2\mathcal{J}_2(t)]t =
\beta[(A_1 + A_2)\mathcal{J}_1(t) + A_2\mathcal{J}_0(t)]t \,.
\end{equation}
Here, the random variables are the net fluxes $\mathcal{J}_{i}(t)$ averaged over the time interval $t$, $i=1,2$ or $i =1,0$. We wrote two alternative divisions of the dissipation flux discussed in figure~\ref{fig2} and we defined
\begin{equation}
\mathcal{J}_0(t) := \mathcal{J}_2(t) - \mathcal{J}_1(t) \,.  
\end{equation}

The detailed fluctuation theorem corresponding to the integral fluctuation theorem (\ref{eq2}) with the second division (\ref{eq3}) is, in the logarithmic form,
\begin{equation}
\label{dft}
\ln\frac{p(j_{1}(t),j_{0}(t))}{p(-j_{1}(t),-j_{0}(t))}
=\beta[(A_{1}+A_{2})j_{1}(t)+A_{2}j_{0}(t)]t \,.
\end{equation}
 $j_i(t)$ in (\ref{dft}), $i=1,0$, denote particular values of random fluxes $\mathcal{J}_{i}(t)$ in (\ref{eq3}). $p$ is the joint probability distribution function of these fluxes and the average in (\ref{eq2}) is performed over this distribution. By definition, the stationary averages of the random fluxes are time independent, $\langle\mathcal{J}_i(t)\rangle = J_i$ for any $t$, but the value of the higher moments of distribution $p(j_{1}(t),j_{0}(t))$ increases with the shortening of time $t$. The convexity of the exponential function provides the second law of thermodynamics
 \begin{equation}
 \label{eq5}
 \langle\sigma\rangle = \beta [(A_1+A_2)J_1 + A_2J_0] t \geq0
 \end{equation}
 to be a consequence of the Jarzynski equality (\ref{eq2}).
 
Let us note that equation~(\ref{dft}) is identical to the fluctuation theorem for the stationary fluxes, earlier introduced by Gaspard \cite{Gasp13} when based on much more complex considerations. But the condition on $t$ is more precisely determined here. The chosen time $t$ must be long enough for the considered ensemble to comprise only stationary fluxes, but finite for observing any fluctuations. All stationary fluxes in the ensemble are statistically independent, hence the probability distribution function $p(j_1(t),j_0(t))$ is the two-dimensional Gaussian~\cite{Sir16}
 \begin{eqnarray}
 \label{gauss}
&& p(j_1,j_0)=\frac{1}{2\pi\mathit{\Delta}_1\mathit{\Delta}_0\sqrt{1-\rho^2}} \\
&&\times  \exp \left\{-\frac{1}{2(1-\rho^2)}\right. 
\left[\frac{(j_1-J_1)^2}{\mathit{\Delta}_1^2}\right. -2\rho\frac{(j_1-J_1)(j_0-J_0)}{\mathit{\Delta}_1\mathit{\Delta}_0}
\left.\left.+\frac{(j_0-J_0)^2}{\mathit{\Delta}_0^2}\right]\right\} ,
\nonumber
 \end{eqnarray}
where the averages $J_i := \langle\mathcal{J}_i\rangle$,  $i = 1, 0$, and the  corresponding variances (the squares of the standard deviations) $\mathit{\Delta}_i^2 := \langle(\mathcal{J}_i - J_i)^2\rangle$ specify the Gaussian marginals for the individual fluxes, and $\rho$ is the correlation coefficient:
\begin{equation}
\rho=\frac{\left<(\mathcal{J}_1-J_1)(\mathcal{J}_0-J_0)\right>}
{\mathit{\Delta}_1\mathit{\Delta}_0} \,.
\label{correlation}
\end{equation}
Here and further on, for the sake of brevity, we omit argument $t$ specifying all the fluxes. The requirement from the Gaussian (\ref{gauss}) to satisfy the detailed fluctuation theorem (\ref{dft}) leads to a system of quadratic equations that link standard deviations $\mathit{\Delta}_1$ and  $\mathit{\Delta}_0$ with averages  $J_1$ and $J_0$:
 \begin{eqnarray}
\mathit{\Delta}_1^2&+&\frac{A_2}{A_1+A_2}\rho
\mathit{\Delta}_0\mathit{\Delta}_1=\frac{2J_1}{\beta(A_1+A_2)t} \,,\nonumber\\ 
 \mathit{\Delta}_0^2& + &\frac{A_1+A_2}{A_2}\rho
\mathit{\Delta}_0\mathit{\Delta}_1
=\frac{2J_0}{\beta A_2t} \,.
 \label{stdev}
 \end{eqnarray}
From (\ref{stdev}) its follows that standard deviations $\mathit{\Delta}_1$ and  $\mathit{\Delta}_0$ are inversely proportional to the square root of $t$. For the lack of correlations, $\rho = 0$, the solutions to (\ref{stdev}) reconstruct our earlier result \cite{Chel17}.

In order to answer the question whether energy dissipation is the only mechanism of entropy production, let us find the fluctuation theorems for fluxes $\mathcal{J}_1$ and $\mathcal{J}_0$ separately. Equations (\ref{stdev}) allow to calculate the ratios of marginals $p(j_1)/p(-j_1)$ and  $p(j_0)/p(-j_0)$, from which it follows that the detailed and integral fluctuation theorem takes the form
\begin{eqnarray}
\label{eq6}
&&\ln\frac{p(j_{1})}{p(-j_{1})} = \beta(A_{1}+A_{2})j_{1}t 
+ \rho \frac{\mathit{\Delta}_0}{\mathit{\Delta}_1} \beta A_{2} j_{1}t
=: \sigma_1(j_1) + \iota_1(j_1) \,, \nonumber\\
&&\langle\exp(-\sigma_1-\iota_1)\rangle = 1 \,,
\hspace{4pc} \langle\sigma_1\rangle +  \langle\iota_1\rangle  \geq 0 \,,
\end{eqnarray}
for the flux $\mathcal{J}_1$ and 
\begin{eqnarray}
\label{eq7}
&&\ln\frac{p(j_{0})}{p(-j_{0})} = \beta A_{2} j_{0}t 
+ \rho \frac{\mathit{\Delta}_1}{\mathit{\Delta}_0} \beta(A_{1}+A_{2}) j_{0}t =: \sigma_0(j_0) + \iota_0(j_0) \,, \nonumber\\
&&\langle\exp(-\sigma_0-\iota_0)\rangle = 1 \,,
\hspace{4pc} \langle\sigma_0\rangle +  \langle\iota_0\rangle  \geq 0 \,.
\end{eqnarray}
for the flux $\mathcal{J}_1$. The averages are performed over the one-dimensional distributions, either $p(j_1)$ or $p(j_0)$. In addition to the functions $\sigma_1$ and $\sigma_0$ that describe separate free energy dissipations, we have obtained additional functions $\iota_1$ and $\iota_0$. The generalized formulations of the second law of thermodynamics with these functions are also listed. As $\mathit\Delta_1$ and $\mathit\Delta_0$ are positive and forces $\beta (A_1+A_2)$ and $\beta A_2$ are of the opposite signs, components $\langle\sigma_i\rangle$ and $\langle\iota_i\rangle$ add up if $\rho$ is negative and subtract if $\rho$ is positive.

The interpretation of functions $\iota_1$ and $\iota_0$ is not obvious. Because the network determining the dynamics of the system does not generally have a bipartite structure unlike the networks for which the concept of Shannon's mutual information was pondered \cite{Saga12,Hart14,Horo14}, the expressions for $\iota_1$ and $\iota_0$ provided by equations~(\ref{eq6}) and (\ref{eq7}) cannot be rewritten using such term. However, these functions actually represent information exchanged between macrodynamics and microdynamics acting as an information reservoir \cite{Mand12,Bara14,Shir16}. We mentioned in subsection 2.2 that the observable action of the machine is its turnovers measured by the net number of the created molecules P$_1$ and P$_2$. Hence, the signal sent by the machine is in the form of two sequences of bits, or rather signs, e.g. $\dots,+,+,-,+,-,+,+,+,-,-,+,\dots$, that describe successive transitions forward or back through the input or output gates, respectively. 

Only the sum of both functions $\iota_1$ and $\iota_0$ is related to mutual information between macrovariables $J_1$ and $J_2$, as from Eqs.~(\ref{dft}), (\ref{eq6}) and (\ref{eq7}), the relationship
\begin{equation}
\label{eq8}
\langle \iota_1 \rangle + \langle \iota_0 \rangle =
- \left\langle \ln\frac{p(\mathcal{J}_1,\mathcal{J}_0)}{p(\mathcal{J}_1)p(\mathcal{J}_0)} 
- \ln\frac{p(-\mathcal{J}_1,-\mathcal{J}_0)}{p(-\mathcal{J}_1)p(-\mathcal{J}_0)} \right\rangle
\end{equation}
results. Let us emphasize that because $\iota_1$ is the function of only $j_1$, and $\iota_0$ is the function of only $j_0$, the value of the difference of the logarithms in (\ref{eq8}) remains the same when averaged over the correlated distribution $p(j_1,j_0)$ or the product $p(j_1)p(j_0)$. Of course, this is not true for the separate components of the logarithms difference in (\ref{eq8}).

Because equation~(\ref{eq8}) is symmetric with respect to the replacement of $\mathcal{J}_1$ with $\mathcal{J}_0$, it does not describe any directed information transfer between both macrodynamics. This means that in addition to the law of free energy conservation in each of the subsystems $i = 1, 0$,  there is an additional law of conservation, implying equality of the information transfered from macrodynamics to microdynamics with the information coming from the environment to macrodynamics as a signal of random passages of the product molecules through the constraints. Indeed, with the help of equations~(\ref{stdev}), equations~(\ref{eq6}) and (\ref{eq7}) can be rewritten directly as
\begin{equation}\label{eff1}
\ln\frac{p(j_{1})}{p(-j_{1})} = \frac{2J_1}{\mathit{\Delta}_1^2} j_1
=: \beta (\tilde{A_1}+\tilde{A_2}) j_1 t
\end{equation}
for the flux $\mathcal{J}_1$ and 
\begin{equation}\label{eff0}
\ln\frac{p(j_{0})}{p(-j_{0})} = \frac{2J_0}{\mathit{\Delta}_0^2} j_0
=: \beta \tilde{A_2} j_1 t
\end{equation}
for the flux $\mathcal{J}_0$. Forces $\tilde{A}_1$ and $\tilde{A}_2$ defined above include, in addition to the forces $A_1$ and $A_2$, corrections to the signals of random passages of the product molecules through the constraints.  These signals play the role of additional constant chemical forces \cite{Shir16} acting on each of the macrodynamics separately. In such approach, the remained macrodynamics together with microdynamics are jointly treated as hidden \cite{Mehl12}.

A comparison of two equivalent formulations of the fluctuation theorem for separate fluxes, (\ref{eq6}) and (\ref{eq7}) with (\ref{eff1}) and (\ref{eff0}), leads to expressions for stochastic informations $\iota_1$ and $\iota_2$ independent of the correlation coefficient $\rho$. Because forces $A_1$ and $A_2$ uniquely determine the values of stationary fluxes $J_1$ and $J_2$ \cite{Kurz14}, only four parameters  $J_i$ and  $\mathit{\Delta}_i$,  $i = 1, 0$, that characterize the Gaussian distribution functions of the separate fluxes, are sufficient to specify four averages: energy dissipations time $\beta D_i := \langle\sigma_i\rangle$ and informations $I_i := \langle\iota_i\rangle$  for $i =1$ and $0$. 

\section{Verification of analytical results in the model system}

\begin{figure}[h]
	\centering
	\includegraphics[scale=0.9]{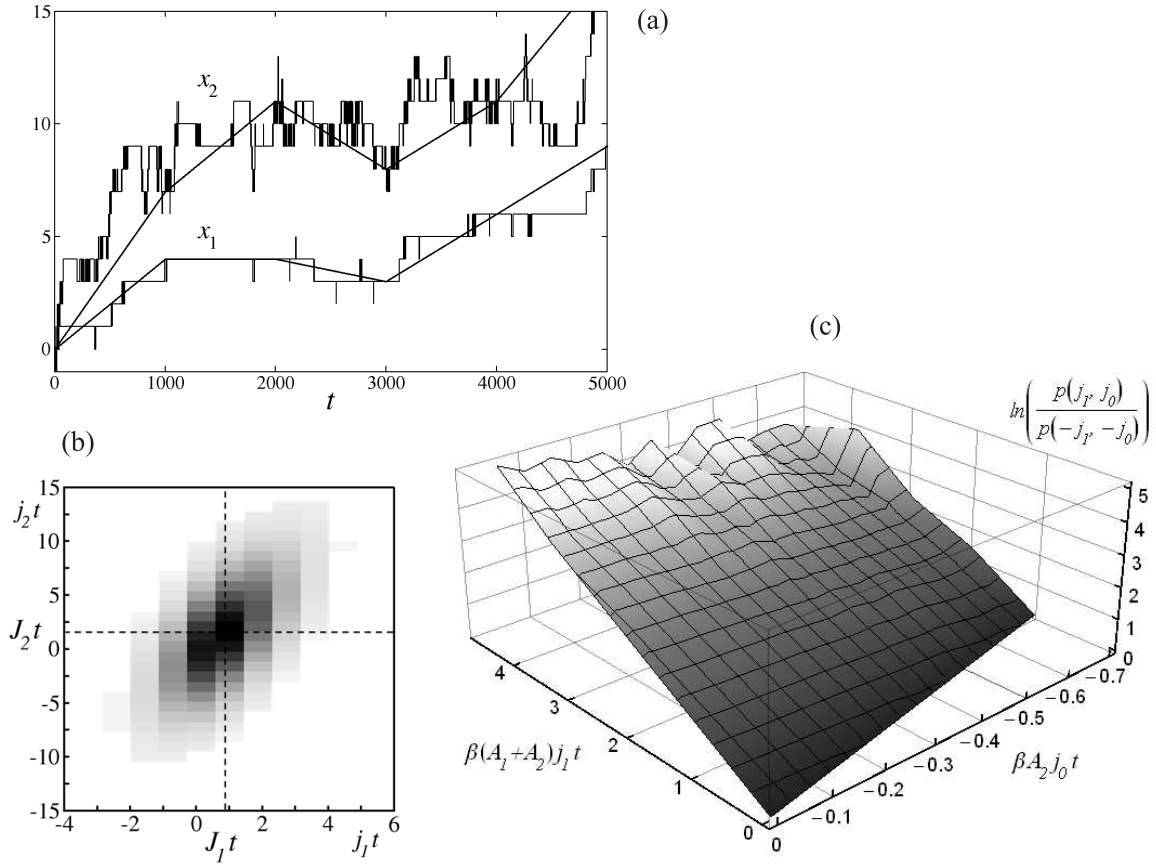}
	\caption{(a)~Simulated time course of the net number of external transitions $x_1$ and $ x_2$ through the input and output gates, respectively, produced in $t$ random walk steps on the network presented in figure~\ref{fig3}(d) with $\beta A_1 = 1$, $\beta A_2 = -0.05$ and the fourfold output gate. Determination of the corresponding fluctuating fluxes $j_1$ and $j_2$ by averaging over successive time periods of $t = 1000$ steps is shown. (b)~The  determined Gaussian distribution of the fluxes $j_1$ and $j_2$. (c)~Numerical verification of the detailed fluctuation theorem (\ref{dft}). Deviations from the plane for large values of $j_1t$ and $j_0t$ result from insufficient statistical material collected in simulations.}
	\label{fig4}
\end{figure}

We performed computer simulations of random walk on the network constructed following the algorithm described in appendix A and shown in figure~\ref{fig3}(d). We assumed $\beta A_1 = 1$ and a few smaller, negative values of $\beta A_2$. The result of each simulation was the time courses of the net numbers $x_1$ and $x_2$ of transitions through the input and output gates, respectively. Exemplary courses are shown in figure~\ref{fig4}(a). Time is counted in the random walk steps.

The thermodynamic description of the nanosystem still required averaging over a finite observation period. To get statistical ensembles of the stationary fluxes, we divided long stochastic trajectories of $10^{10}$ random walk steps into segments of equal lengths $t$. For the fluxes $j_i$ to be statistically independent, the selected time $t$ must have been longer than the mean duration of the transient stage completing the free energy-transduction cycle. In appendix B we prove that, although the microscopic dynamics is a Markov process, the time courses $x_1$ and $x_2$ are not Markov processes but continuous time random walks \cite{Metz00,Metz14}. We also show that for the assumed conditions of motion, the reasonable time of averaging is $t = 1000$ random walk steps. The result of averaging is pairs of fluxes $j_i(t) = \Delta x_i(t)/t$,  $i = 1, 2$, which is depicted also in figure \ref{fig4}(a).

The distribution of random fluxes $j_1(t)$ and $j_2(t)$ found for $\beta A_1 = 1$, $\beta A_2 = -0.05$ and the fourfold output gate is depicted in figure~\ref{fig4}(b). The distribution is actually Gaussian. The thermodynamic behavior of a single biological nanomachine is simple diffusion in a two-dimensional parabolic potential on the plane $(j_1, j_2)$ with a minimum at point $(J_1, J_2)$. After exchanging the variables from $j_1$ and $j_2$ to $j_1$ and $j_0$, we get the two-dimensional distribution $p(j_1,j_0)$. This distribution actually satisfies fluctuation theorem (\ref{dft}), as illustrated in figure~\ref{fig4}(c).

\begin{figure}[h]
	\centering	
	\includegraphics[scale=0.85]{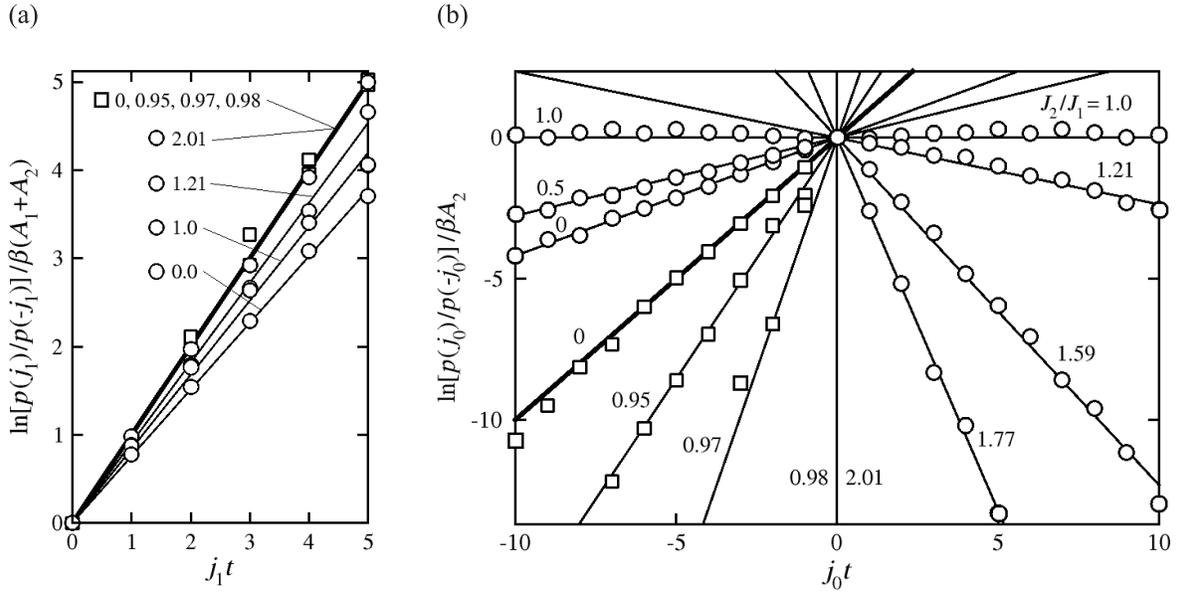}
	\caption{The generalized fluctuation theorem dependences found in the random walk simulations on the network shown in figure~\ref{fig3}(d) with the single output gate (the squares) and the fourfold output gate (the circles). We assumed $\beta A_1 = 1$ and a few smaller, negative values of $\beta A_2$ determining the ratio of averaged fluxes $\epsilon = J_2/J_1$ noted in the graphs. (a)~The case of marginal $p(j_1)$, compare equation~(\ref{eq6}). (b)~The case of marginal $p(j_0)$, compare equation~(\ref{eq7}). In order to clearly distinguish the contributions from dissipation and information, the results were divided by dimensionless forces $\beta (A_1 + A_2)$ and $\beta A_2$ for $i =$ 1 and 0, respectively. In this way, the bold lines in both graphs, with unit tangent of inclination angle, correspond to the dissipation components. Let us recall that $\beta (A_1+A_2)$ is positive while $\beta A_2$ is negative.} 
	\label{fig5}
\end{figure}

From the two-dimensional distributions, we calculated the marginal distributions $p(j_1)$ and $p(j_0)$ for all simulated trajectories. The logarithms of the ratio of marginals $p(j_i)/p(-j_i)$ are presented in figures~\ref{fig5}(a) and (b) as the functions of $j_i\,t$ for $i = 1$ and $0$, respectively. All the dependences are actually linear as predicted by generalized fluctuation theorems (\ref{eq6}) and (\ref{eq7}). 

For the single output gate (the squares in figure~\ref{fig5}), we got both $\iota_1$ and $\iota_0$ positive, of the same sign as $\sigma_1$ and $\sigma_0$. This is consistent with the determined negative values of correlation coefficient $\rho$. Both information contributions differ from zero only for $\epsilon$ close to unity, i.e. for very small values of force $A_2$, which does not break practically the detailed balance condition for the transitions through the only output gate. In consequence, information $\iota_1$ is unnoticeable in figure~\ref{fig5}(a) ($A_1 + A_2 \approx A_1$), and the signal of the passages through the output gate, determining information $\iota_2$ in figure~\ref{fig5}(b), is a usual white noise. For the fourfold output gate (the circles in figure~\ref{fig5}), the situation is much more interesting. Now, information contributions $\iota_i$ are of the opposite sign to dissipation contributions $\sigma_i$, which is consistent with the positive values of correlation coefficient $\rho$.

\section{Energy processing versus information processing in entropy production}

\begin{figure}[ht]
	\vspace{1pc}	
	\centering
	\includegraphics[scale=0.9]{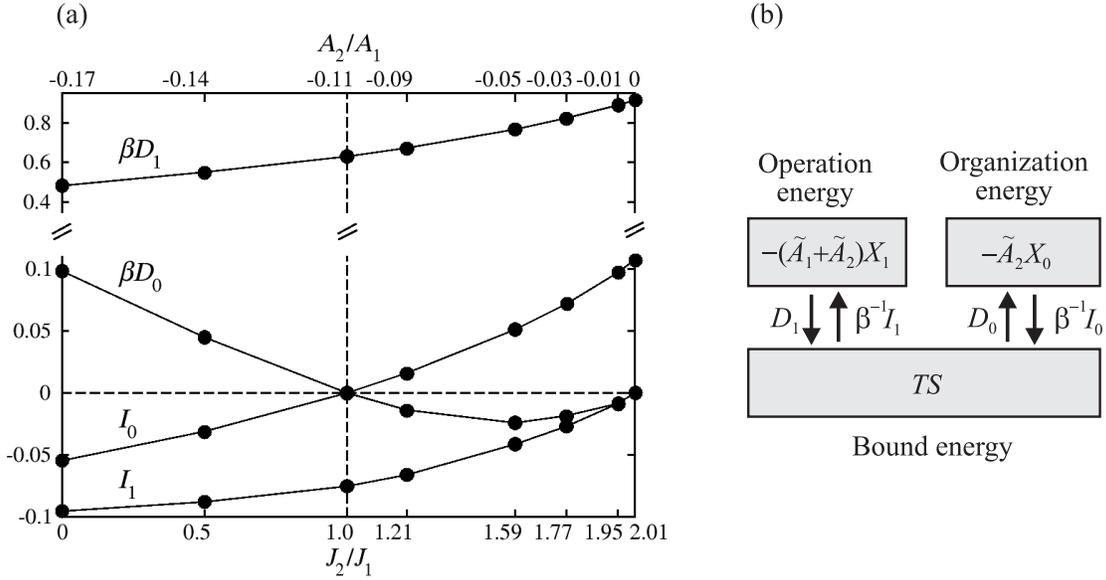}
	\caption{(a)~The contribution of dissipation $\beta D_i$ and information $I_i$ to entropy production for the model system under study with the fourfold output gate, presented as function of the ratio of fluxes $J_2/J_1$ or, equivalently, the forces $A_2/A_1$. $ i = 1 $ in the case of operation energy processing, and $ i = 0$ in the case of processing of organization energy. All quantities are counted in nats (natural logarithm is used instead of binary logarithm) per random walk step. (b)~A general scheme of communication between macrodynamics, which determines free energy and microdynamics, which determines bound energy for $J_2/J_1 > 1$. $\tilde{A_1}$ and $\tilde{A_2}$ denote the effective forces that take additionally into account the signals coming from the environment.}
	\label{fig6}
\end{figure}

For the fourfold output gate, the averaged energy dissipations $\beta D_i := \langle\sigma_i\rangle$ and informations $I_i := \langle\iota_i\rangle$  for $i =1$ and $0$ are presented in figure~\ref{fig6}(a) as the functions of the degree of coupling $\epsilon = J_2/J_1$. The values determined from the inclination of the straight lines in figure~\ref{fig5} coincide with the values determined directly by four parameters: $J_i$ and  $\mathit{\Delta}_i$,  $i = 1, 0$, what predicts the comparison of equations (\ref{eq6}) and (\ref{eq7}) with (\ref{eff1}) and (\ref{eff0}).  

It is clearly seen that, actually, information $ I_i $  is always of the opposite sign to dissipation $\beta D_i $. For $J_2/J_1 < 1$, which corresponds to large values of output force $\beta A_2$, information $I_0$ is negative, i.e. taken from the microdynamics, to be next subtracted from the positive dissipation $\beta D_0$. For $J_2/J_1 > 1$, which corresponds to small values of output force $\beta A_2$, information $I_0$ is positive, i.e. sent to the microdynamics and the production of entropy $\beta D_0$ becomes negative. The biological molecular machines, for which this is the case, may be said to really act as Maxwell's demons, although they do not use the information creation directly for work, but only for reduction of energy losses.

Our general approach to the describtion of the thermodynamics of nanoscopic machines is based on the division of free energy into the operation and organization energy, and the introduction of the effective thermodynamic forces that take into account the signals encoded in fluctuations. The effective forces in the chemo-chemical studied nanomachines are described by equations  (\ref{eff1}) and (\ref{eff0}), and are noted in free energy division shown in figure~\ref{fig6}(b). Such generalized free energy exchanges both dissipation and information with thermodynamic entropy that plays the role of memory thanks to a transient breaking of ergodicity.

If the nanomachine has no choice possibility, both $D_1$ and $\beta^{-1} I_1$, as well as $D_0$ and $\beta^{-1} I_0$, are positive and information or, rather, noise processing results in an additional energy loss. Only in the nanomachines that choose randomly the way of doing work and transmit information about it to microdynamics, $D_0$ and $\beta^{-1} I_0$ are of the opposite sign, which makes $D_0$ to be negative, when $\beta^{-1} I_0$ is positive. However, according to the inequalities specified in (\ref{eq6}) and (\ref{eq7}), the resultant entropy productions on both subsystems must always be non-negative: 
\begin{equation}
\label{eq9}
D_1 + \beta^{-1}I_1 \geq 0 \,, ~~~~~~~~D_0 + \beta^{-1}I_0 \geq 0 \,.
\end{equation}

\section{Discussion}

Under physiological conditions, the protein molecular machines fluctuate constantly between lots of conformational substates composing their native state. The probabilities of visiting individual substates are far from the equilibrium and determined by the concentration of the surrounding molecules involved in the process. During the full cycle of free energy-transduction, the possibility to choose different realizations of the free energy-accepting reaction results in the transient limitation of the dynamics to different regions of the conformational network, that is, to breaking the ergodicity \cite{Engl06,Metz14}. The transient ergodicity breaking transforms the machine's internal dynamics into a memory for storing and manipulating information. Information is erased each time the energy-donating reaction starts the next cycle from the input gate. The storage capacity of memory is higher the larger and more complex is the network of conformational substates. This network is particularly large in the case of protein motors, since it then contains the substates of the motor bound to the whole track on which it moves \cite{Kurz06}, compare figure~\ref{fig3}(b).

Information is exchanged between memory and the thermodynamic variables. Two components of free energy determine the thermodynamic state of the machine. The first, proportional to the concentration of the input reaction product molecules $X_1$, we refer to as the operation energy.  And the second, proportional to the difference in the concentrations of the output and input reaction product molecules $X_2-X_1$, we refer to as the organization energy. Similarly as work and dissipation, information is also a change in these well defined functions of state. In the protein machines with no possibility of choice, changes in organization are related, like in the macroscopic machines, only to energy losses. Otherwise, in the protein machines with such a possibility, changes in organization can essentially reduce these losses.

In figure~\ref{fig6}(a), which presents the most important result of the paper, two cases of zero dissipation of energy $D_0$ stand out. The first one is realized for tight coupling, $\epsilon = J_2/J_1 = 1$, i.e. $J_0 = 0$, and occurs simultaneously with the zero transfer of information $I_0$. The necessary condition for it is a special, critical value of output force $A_2$, distinguished in figure~\ref{fig6}(a) by the dashed vertical line. There are some arguments that, effectively, such the force value is realized for processive motors with a feedback control between their two identical components \cite{Yild08,Bier11}. The information transfer to memory takes place only for the force of a value less than critical. A chosen monomer collects this information from the second monomer (the ``measurement'') and exerts on this monomer some force (the ``control''). This force, when added to the original force, gives the resultant force of the critical value. The whole resultant critical force is exerted on the second monomer, which does not exchange information at all and works only as a tightly coupled perfect machine. The process is repeated with the conversion of one monomer to the other, so that the external forces act alternately, as in scheme \ref{fig3}(e). Consequently, the network of conformational transitions for the pair has a bipartite structure \cite{Horo14}. The suggested mechanism seems to be in harmony with the consensus kinesin-1 chemo-mechanical cycle \cite{Hanc16} and, in a broader perspective, it could help answer the certainly interesting question: why do most protein machines operate as dimers or higher organized assemblies?

The second case of zero dissipation $D_0$ takes place for $\beta A_2$ tending to 0, when the system is no longer working as a machine, but dissipation $D_1$ is used for collecting information $I_0$. Such conditions are met by the transcription factors in their one-dimensional search for the target on DNA, intermittent by three-dimensional flights \cite{Kolo11}. Following our theory, the flights correspond to the randomly chosen external transitions, hence the one-dimensional sliding between them is not  simple diffusion but the active continuous time random walk (compare appendix B). In fact, the external transitions, assumed in the present model to be very fast, are also continuous time random walk \cite{Lium17}. During information collection, ergodicity is broken, which agrees with the suggestion that the macromolecule of the transcription factor is derived from the equilibrium from the very beginning of the search \cite{Kolo11}.

At the end, let us take the liberty for a few general remarks. The generalized second laws of thermodynamics (\ref{eq9}) were justified only for isothermal processes in the nanoscopic machines. Their universality remains an open problem. One thing is certain. In order for information and energy dissipation to be of opposite signs, two necessary conditions must be met. These are the presence of an intermediate level of stochastic dynamics and the possibility of a choice. Two examples of systems having such properties are well known. The first one is thermodynamic systems occurring in the conditions of two phases coexistence, whose organization is specified by an extra thermodynamic variable that survived stochastization \cite{Penr79}, referred to as the order parameter \cite{Call85} or, in various contexts, the emergent \cite{Ande72} or structural \cite{Kurz06} variable. Here, the random variable is the local order parameter. The transient non-ergodic state of dynamics may last short, as in the case of the fogg condensation in a gas-liquid mixture, or very long time, as in the case of glass a liquid-solid mixture. The second example is quantum systems occurring in states entangled with the environment, whose organization  is determined by an extra variable that survived decoherence, identified with the classical variable \cite{Zure03,Zure09}. Here, the value of this variables found in the local measurement is random. In both examples, the random choice, which is the source of information, should be considered in the context of energy dissipation.

\section*{Appendix A. Specification of the computer model}

The algorithm of constructing the stochastic scale-free fractal trees is from Goh et al. \cite{Goh06}. Shortcuts, though more widely distributed, were considered by Rozenfeld, Song and Makse \cite{Roze10}. Here, we randomly chose three shortcuts from the set of all the pairs of nodes distanced by six links. The network of 100 nodes in figure~\ref{fig3}(d) is too small to determine its scaling properties, but a similar procedure of construction applied to $10^5$ nodes results in a scale-free network, which is fractal on a small length-scale and a small world on a large length-scale. 

To provide the network with stochastic dynamics, we assumed the probability of changing a node to any of its neighbors to be the same in each random walking step \cite{Kurz14,Chel17}. Then, following the detailed balance condition, the microscopic free energy of a given node is proportional to the number of its links (the node degree). The most stable nodes are the hubs, which are the only practically observed conformational substates under equilibrium conditions. The assumed dynamics is independent of temperature, and therefore not very realistic, but this problem is not the subject of our research here. For given node $l$, the transition probability to any of the neighboring nodes per one random walking step is one over the number of links and equals $(p_l^{\rm eq}\tau_{\rm int})^{-1}$, where $p_l^{\rm eq}$ is the equilibrium occupation probability of the given node and $\tau_{\rm int}$ is the mean time to repeat a chosen internal transition, counted in the random walking steps. This time is determined by the doubled number of links minus one \cite{Chel17}, $\tau_{\rm int}=2\dot(100+3-1)=204$ random walking steps for the 100 node tree network with 3 shortcuts assumed.

The forward external transition probability, related to stationary concentration [P$_i$], is determined by the mean time of external transition $\tau_{\rm ext}$, and equals $(p_i^{\rm eq}\tau_{\rm ext})^{-1}$ per random walking step, $p_i^{\rm eq}$ denoting the equilibrium occupation probability of the initial input or output node ($i=1,2$, respectively). The corresponding backward external transition probability is modified by detailed balance breaking factors $\exp (-\beta A_i)$. The assumed mean time of forward external transition $\tau_{\rm ext}=20$ was one order of the magnitude shorter than the mean time of internal transition $\tau_{\rm int}=204$, so that the whole process is controlled by the internal dynamics of the system. The opposite situation, when the process is controlled by the external spatial diffusion of the involved molecules, is not considered in the present paper.

\section*{Appendix B. Determination of the flux averaging time}

\begin{figure}[h!]
	\centering
	\includegraphics[scale=1.75]{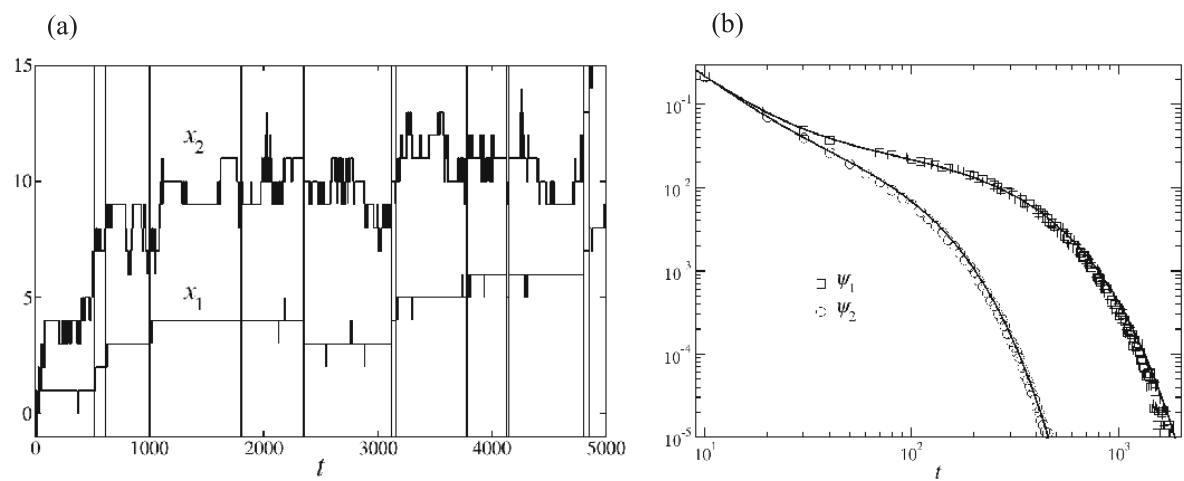}
	\caption{(a)~Simulated time course of the net number of external transitions $x_1$ and $ x_2$ through the input and output gates, respectively, produced in $t$ random walk steps. Vertical lines determine the dwell times for subsequent external transitions through the input gate, neglecting periods of less than 10 steps. (b)~Log-log plot of distributions $\psi_i(t)$ between the successive external transitions through	a single input gate ($i=1$) and a fourfold output gate ($i=2$). }
	\label{fig7}
\end{figure}

The result of a random walk simulation on the network we described is a time course
of the net numbers $x_1$ and $x_2$ of external transitions through the input and output gate, respectively. Figure \ref{fig7}(a), in which time $t$ is counted in random walk steps, shows an example of such a course for $\beta A_1 = 1$,  $\beta A_2 = -0.05$ and the fourfold output gate. The simulated time course determine, above all, the dwell times for the subsequent external transititions. However, it needs to point out numerous vertical dashes of imperceptible width in the depicted courses. In the real experiments such transitions are not observed due to inertia and a finite time resolution of the instrument. We also used a smoothing procedure for all time courses with a resolution of 10 steps.  In figure~\ref{fig7}(b), the dwell time distributions $\psi_1(t)$ and $\psi_2(t)$ are shown for the successive transititions through the input and output gates, respectively, determined from the smoothed long stochastic courses of $10^{10}$ random walk steps, the beginning of which is shown in  figure~\ref{fig7}(a).

Both distributions $\psi_1(t)$ and $\psi_2(t)$ fit very well the dependences~\cite{Kurz98}
\begin{equation}
\psi_i(t) = \left[ (1-c_i)(\eta_it)^{-\alpha_i}/\sqrt{\pi} + c_i \right] {\rm e}^{-t/\tau_i} \,.
\label{fit}
\end{equation}
They include a certain power law range before the exponential completion,
which results from fractality of the network and  indicates that the stochastic processes $x_i(t), i = 1, 2$, are not the Markov processes, but rather a complex continuous time random walks \cite{Metz00}. Distribution $\psi_1(t)$ determines durations of the free energy-transduction cycles. From the fitting, we found $\eta_1=0.18$, which approximates the reciprocal of the smoothing time, and $\alpha_1=1.8$, which is an exponent that specifies the fractal properties of the tree-like network. The contribution from the exponential decay $c_1=0.03$ is very small, but it is the long mean time of its decay $\tau_1=230$ that determines duration of the cycle.  For comparison, we found the mean-first passage time between the most distant nodes of the network to be equal 710 random walking steps. Taking all of it into account, we decided to choose a reasonable flux averaging time $t = 1000$ computer steps in the case of the network equipped with the fourfold output gate.

\section*{Acknowledgements}

M\,K thanks Ya\c{s}ar Demirel and Herv\'{e} Cailleau for discussing the problem in the early stages of the investigation.

\section*{Author contributions statement}

The general concept and the theory is mainly due to M\,K, who also wrote the manuscript. The specification of the critical branching tree model and the numerical simulations are mainly due to P\,C.

\end{document}